# Single-Atom Photocatalysts on $TiO_2$ : Insights from X-ray Absorption Spectroscopy

Yingying Li[a], David Morris[b], Peng Zhang[*b]

[a] State Key Laboratory of Integrated Optoelectronics, Key Laboratory of UV Light-Emitting Materials and Technology of Ministry of Education, School of Physics, Northeast Normal University, Changchun, 130024, China.

[b] Department of Chemistry, Dalhousie University, Halifax, NS, Canada.

Corresponding Author:peng.zhang@dal.ca

Abstract

Surface modification of $TiO_2$ with single-atom catalysts (SACs) is an effective strategy for enhancing photocatalytic efficiency. However, thorough characterization of SACs at the atomic scale remains challenging. X-ray absorption spectroscopy (XAS) offers unique advantages for the in-depth analysis of $TiO_2$-supported SACs. By employing XAS, the local atomic structure, oxidation state, and electronic properties of the SACs, as well as the underlying photocatalytic mechanism, can be revealed. Herein, we present a short review on the application of XAS in studying $TiO_2$-supported SACs. We first elucidate the key role of XAS in simultaneously probing the structure and electronic properties of monometallic SACs across different periods. Next, we discuss XAS studies of bimetallic SACs from the perspective of each constituent element and highlight the element-specific capabilities of XAS for analyzing multi-element SACs. Finally, we demonstrate how *in situ* XAS can effectively monitor structural and electronic property changes in SACs under real photocatalytic reaction conditions. Overall, this review highlights the unique advantages of XAS in achieving a more comprehensive understanding of the structure−property relationships in SACs, ultimately aiding the rational design of future photocatalysts. Additionally, we provide practical suggestions for utilizing XAS more efficiently in the analysis of various SAC systems.

1. Introduction

TiO$_2$, a typical semiconductor material, has been widely used in photocatalysis and remains a subject of ongoing research.[1] To further improve its photocatalytic efficiency, various surface modification strategies have been proposed,[2] including heterojunction construction and cocatalyst modification.[3] Among these, single-atom catalyst (SAC) modified TiO$_2$ exhibits excellent photocatalytic performance.[4] SACs maximize atomic utilization and significantly enhance catalytic activity and selectivity through unique interfacial interactions with the $TiO_2$ support.[5] Compared to nanoparticles, SACs composed of elements such as Pt, Au, Pd, Rh, and Cu demonstrate superior photocatalytic activity when atomically dispersed.[6] However, fully elucidating their atomic-scale structure and electronic properties remains a significant challenge.

X-ray absorption spectroscopy (XAS) has emerged as a powerful technique to analyze the atomic structure and electronic properties of SACs.[6d,7] Traditionally, high-angle annular dark-field scanning transmission electron microscopy (HAADF-STEM) is used to provide atomic-scale structural information on SACs, while X-ray photoelectron spectroscopy (XPS) is used to reveal their electronic properties.[8] XAS allows for both atomic-scale structural information and electronic properties to be analyzed using one technique, making it a critical tool in SAC research due to its element specificity, high sensitivity, and applicability under diverse support and reaction conditions.[9] Specifically, X-ray absorption near-edge structure (XANES) primarily elucidates electronic properties and charge transfer mechanisms in catalytic processes, while extended X-ray absorption fine structure (EXAFS) provides detailed atomic structural information.[10] These two spectral regions are complementary, for instance, changes in the Ti-O coordination number (CN) observed in EXAFS often correlate with variations in the Ti oxidation state or electronic structure detected in the

XANES region. [11]

For TiO$_2$-supported SACs, XAS provides crucial information, including oxidation states, d-electron density, CNs, and bond distances.[12] Notably, for transition metals, XAS serves as an integrated technique for characterizing both electronic and structural properties.[13] Additionally, it can identify light atoms such as O and Cl in the coordination environment.[14] For metals with similar atomic numbers, which are difficult to resolve using imagining techniques, XAS is capable of distinguishing them based on their distinct absorption edge energies.[15] Furthermore, XAS can be performed under atmospheric pressure in air, as high-energy X-rays exhibit strong penetration ability and are not significantly absorbed by reaction gases or air.[16] By adjusting the test temperature and atmosphere, changes in the electronic and structural properties of metal SACs can be monitored in real time, revealing the dynamic evolution of photocatalytic activity.[17]

Herein, we present a short review on the application of XAS in the analysis of TiO$_2$-supported SACs. First, we summarize the advantages of XAS for characterizing both monometallic and bimetallic SACs. Subsequently, we highlight the critical role of *in situ* XAS in studying photocatalytic reactions. This review aims to provide critical perspectives and practical guidance for the XAS analysis of TiO$_2$-supported SACs, with the goal of advancing the understanding of their photocatalytic mechanisms and enabling the rational design of future photocatalysts.

2. XAS of TiO$_2$-supported monometallic SACs

2.1 5d-transition metal SACs

For 5d-transition metals, imaging techniques such as HAADF-STEM have been utilized to achieve atomic-scale structural analysis.[18] However, these techniques are inherently limited in their ability to provide electronic information.[19] XAS has emerged as an integrated analytical technique capable of simultaneously characterizing both electronic and structural properties. Specifically, the L$_3$-edge excitation corresponding to the $2p_{3/2}$ to $5d$ transition under X-ray irradiation provides valuable information about the electronic properties.[20] XANES reveals the oxidation states of metals by probing d-electron density, which can help identify the optimal valence state for catalytic activity.[21] EXAFS analysis provides detailed atomic-scale structural information, enabling the probing of interactions at the metal-TiO$_2$ interface.[22] These insights are crucial for exploring catalytic mechanisms and guiding the development of high-performance photocatalysts.

Chen et al. successfully assembled single Pt atoms on a defective TiO$_2$ support for photocatalytic hydrogen production.[23] From the XANES spectra of the Pt L$_3$-edge (Figure 1a), the white-line intensity of Pt$_1$/TiO$_2$ falls between that of Pt foil and PtO$_2$, indicating charge transfer between single Pt atoms and surrounding oxygen atoms, arising from enhanced metal–support interactions. FT-EXAFS analysis (Figure 1b-1d) confirmed the isolated nature of the single Pt atoms. Based on the CN and bond distances derived from EXAFS fitting, an atomic model was constructed with the aid of density functional theory (DFT) calculations. The valence states of O and Ti were further examined by XPS, supporting the interpretation of the electronic state of the SACs.

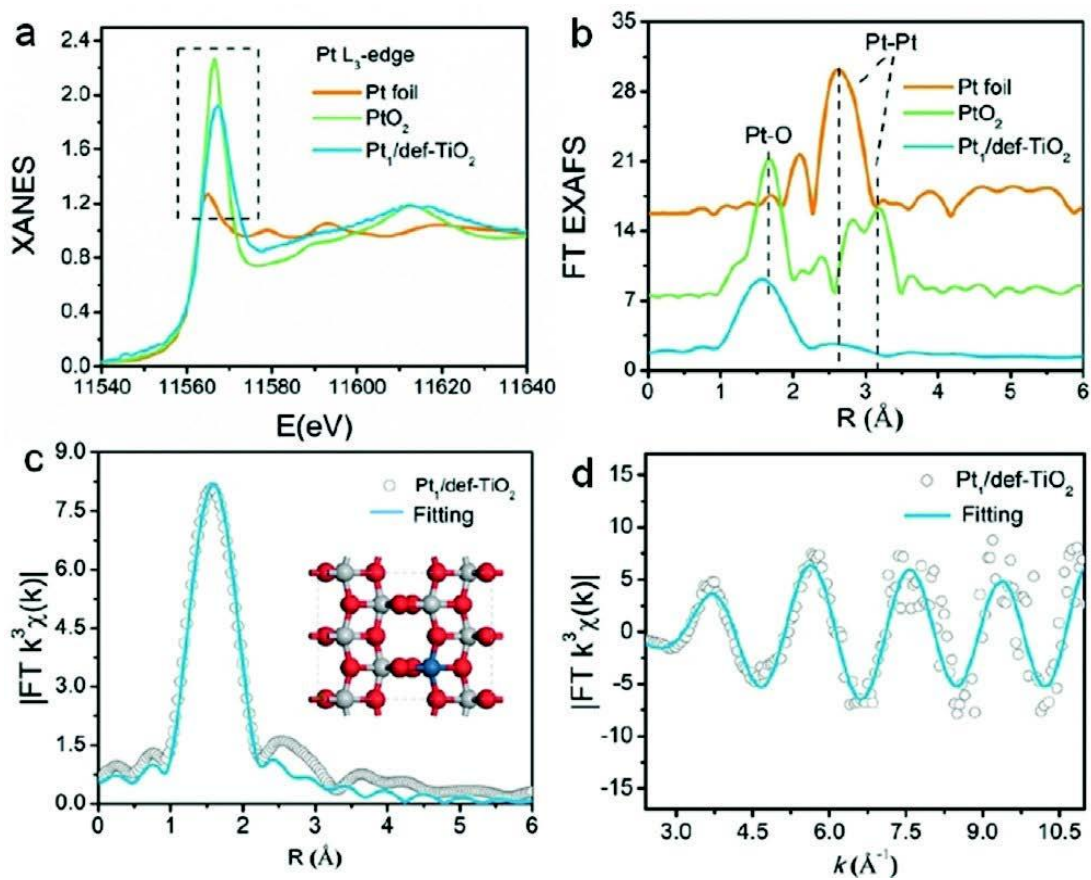

Figure 1: (a) XANES spectra of the Pt $L_3$-edge for the $Pt_1/TiO_2$ catalyst, Pt foil, and $PtO_2$; (b) Fourier-transformed (FT) EXAFS spectra of $Pt_1/TiO_2$ catalyst, Pt foil, and $PtO_2$. (c) EXAFS R-space fitting curve (blue line) and experimental data (circles) for the $Pt_1/TiO_2$ catalyst. (d) EXAFS k-space fitting curve (blue line) and the experimental data (circles) for the $Pt_1/TiO_2$ catalyst.[23]

Feng et al. reported the uniform dispersion of single tungsten atoms on the surface of $TiO_2$ nanoparticles, forming isolated active sites and thereby significantly enhancing $CO_2$ reduction efficiency.[24] It is worth noting that distinguishing W and Ti atoms based on their brightness in HAADF-STEM is particularly challenging, as shown in Figure 2a. XAS provided the most direct evidence confirming the existence of single W atoms in this work (Figure 2b), underscoring its unique advantages over imaging techniques for atomic-scale structural characterization. Similar studies have also used XAS to probe the atomic structures of W-doped $TiO_2$, due to the large difference in their absorption-edge energies.[25] Collectively, these findings highlight the indispensable role of XAS in elucidating the structural and electronic properties of SACs.

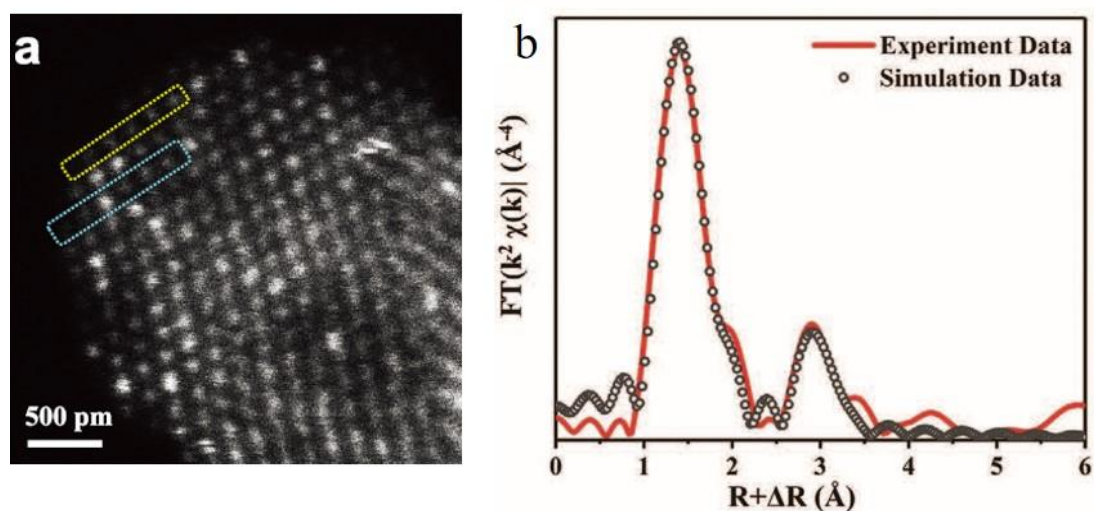

Figure 2: (a) HAADF-STEM image of $W_1/TiO_2$. (b) Comparison of the FT-EXAFS spectra between the experimental data and the fitted results for $W_1/TiO_2$.[24]

Overall, for $TiO_2$ structures modified with heavy single atoms, it is necessary to combine HAADF and EXAFS to resolve the structure of SACs on the $TiO_2$ surface. To elucidate the electronic properties, XANES and XPS are valuable analytical tools.[26] XAS can simultaneously provide electronic and structural information, making it a powerful technique for the comprehensive characterization of such catalytic systems.

2.2 4d-transition metal SACs

HAADF-STEM has been employed for imaging heavier metal elements at the atomic scale.[27] However, this technique is limited in its ability to resolve light elements surrounding metal atoms, which play a critical role in photocatalytic activity.[28] In contrast, XAS provides crucial insights into atomic structure, particularly the chemical bonds between metal atoms and light elements such as Cl and O. [29]

Liu et al. designed an atomically dispersed palladium–titanium oxide catalyst ($Pd_1/TiO_2$) with a Pd content of up to 1.5%, which showed high catalytic activity for C=C bond hydrogenation, achieving nine-fold higher performance than commercial Pd catalysts.[14] EXAFS analysis revealed that only Pd-O bonds were observed at 1-2 Å range, with no Pd-Pd bond peak, as shown in Figure 3a. Notably, the reaction rate of $Pd_1/TiO_2$ was maintained after 20 cycles, whereas that of unsupported $H_2PdCl_4$ decreased after the second cycle. They found that the presence of Pd-Cl bonds (Figure 3b) hindered catalysis, highlighting the crucial role of light-element coordination around metal atoms in directly affecting catalytic activity. XAS plays a key role in distinguishing these light elements, a task unachievable through imaging techniques alone.

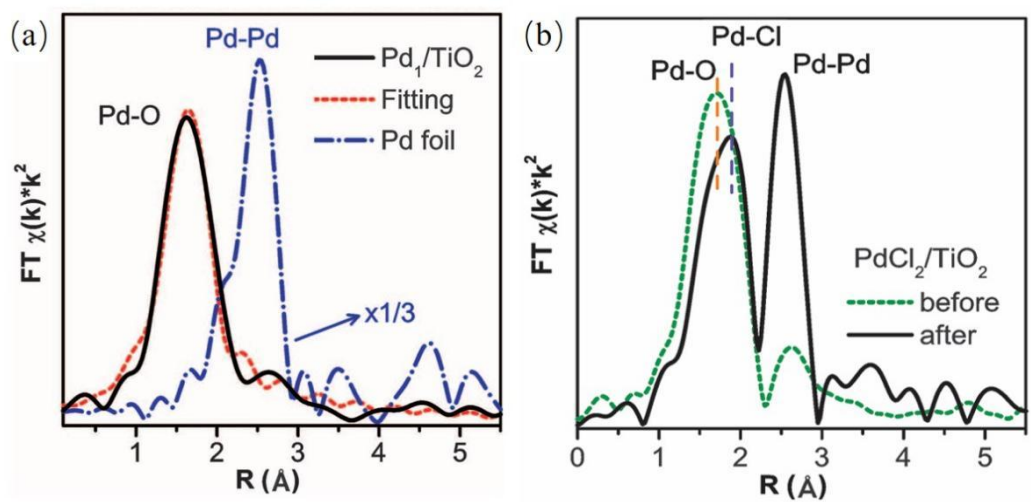

Figure 3: (a) FT-EXAFS spectra of $Pd_1/TiO_2$ and bulk Pd foil at the Pd K-edge, showing the coordination environment of Pd atoms; (b) FT-EXAFS spectra at the Pd K-edge of $PdCl_2/TiO_2$ before and after the catalytic reaction.[14]

Guo et al. employed XANES to analyze the electronic structure and oxidation state of Pd in $Pd/TiO_2$ catalysts.[30] By comparing the spectra with PdO and Pd foil references, they confirmed that the Pd SACs exhibited a positive valence state, consistent with the XPS results. In this study, XANES was used not only for structural characterization but also to reveal the catalytic reaction mechanism, thereby aiding performance optimization.

For 4d transition metals, research often focuses on K-edge analysis, which corresponds to the $1s$ to $4p$ electronic transition.[31] However, the role of d-electrons is more important in catalytic processes.[32] The L-edge absorption ($2p \rightarrow 4d$) in the XAS spectrum is therefore more valuable for revealing catalytic mechanisms.[33] By combining K-edge and L-edge information, a more comprehensive understanding of the electronic properties of 4d transition-metal SACs can be achieved.[34] However, the relatively low absorption energies of the L-edge necessitate the use of soft X-ray.[35] When designing soft X-ray absorption spectroscopy experiments, it is essential to identify and select a synchrotron beamline that covers the required photon energy range of the absorption edge. Representative soft X-ray beamlines are listed in Table 1.

Table 1: Representative soft X-ray beamlines at synchrotron light sources.

| Light Source | Beamline | Energy Range |
| --- | --- | --- |
| Shanghai Synchrotron Radiation Facility (SSRF) | BL14W1 | 4.5-50KeV |
| High Energy Photon Source (HEPS) | ID46 | 4.8-45keV |
| Beijing Synchrotron Radiation Facility (BSRF) | 1W1B | 4.8-22.8keV |
| Beijing Synchrotron Radiation Facility (BSRF) | 4B7B | 50eV-1700eV (XANES) |
| Canadian Light Source (CLS) | SXRMB | 1.7-10 KeV |
| Advanced Photon Source (APS) | 9-BM | 2.1-40 KeV |
| The European Synchrotron (ESRF-EBS) | ID03 | 11.0-60.0 KeV |
| MAX IV | FlexPES | 43 – 1550 eV |
| Spring-8 | BL27SU | 2.1-3.3 KeV |

2.3 3d-transition metal SACs

For $TiO_2$-supported metals, 3d transition metals such as Cu, Fe, Co and Ni, which are in the same period as Ti, are considered promising alternatives to noble metals due to their variable valence and cost-effectiveness.[36] However, for elements within the same period, it is challenging to distinguish them using imagine techniques. XAS overcomes this limitation due to the unique absorption edges of each element. For example, it is difficult to distinguish between Cu and Ti in HAADF-STEM images, as shown in Figure 4a. However, the Cu K-edge is ~8979 eV, whereas the Ti K-edge is ~4966 eV. [37] This pronounced energy difference allows these two elements to be clearly distinguished by XAS.

Fang et al. employed EXAFS to investigate the coordination environment of Cu on $TiO_2$-supported catalysts, as shown in Figure 4b. [38] For $Cu_1/TiO_2$ SACs, the main EXAFS peak corresponded to the Cu-O interatomic distance, while the second shell indicated Cu-Ti interactions, confirming the formation of a Ti-O-Cu structure. For $CuO_x/TiO_2$ nanoparticles, EXAFS analysis suggested coexistence of CuO and $Cu_2O$ species. XANES elucidated the oxidation state and coordination environment of isolated Cu atoms, as shown in Figure 4c. In $Cu_1/TiO_2$ SACs, the pre-edge features of the Cu XANES aligned with those of the $Cu_2O$ reference, indicating that the Cu single atoms primarily exist in the $Cu^{+1}$ oxidation state.

In many cases, a variety of characterization techniques are used to identify the valence state of 3d SACs.[39] For example, Shen et al. confirmed that the valence state of Cu was between +1 and +2 through

XANES.[40] The intermediate valence state was further confirmed by Cu LMM Auger electron spectra, XPS analysis and theoretical calculations. This approach underscores the importance of using multiple complimentary techniques for accurate electronic structure determination.

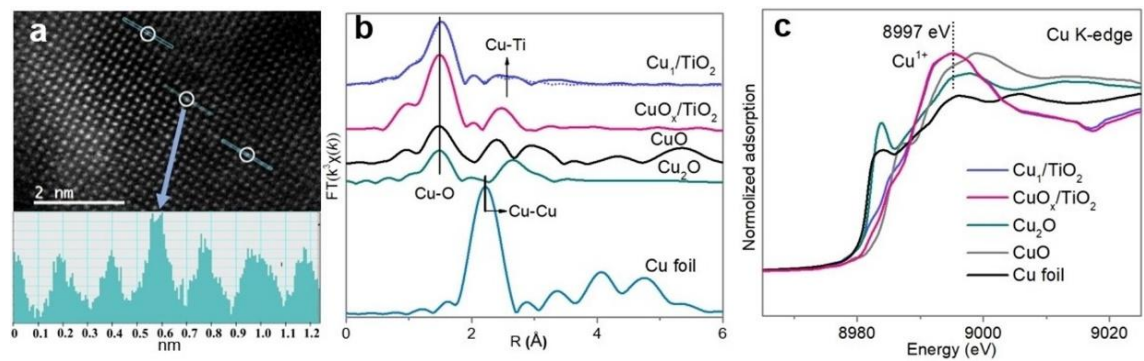

Figure 4: (a) HAADF-STEM image of $Cu_1/TiO_2$ SACs; (b) Normalized FT_EXAFS spectra of the Cu K-edge for $Cu_1/TiO_2$ and $CuO_x/TiO_2$ catalysts; (c) XANES spectra of the Cu K-edge for $Cu_1/TiO_2$ SAC and $CuO_x/TiO_2$ catalyst.[38]

3. XAS of $TiO_2$ supported bi-metallic SACs

3.1 Bi-metallic elements from different periods

XAS is particularly powerful for analyzing the electronic and structural information of individual elements in multi-metallic systems. By combing XANES and EXAFS data, detailed information on each constituent element can be obtained. Furthermore, by integrating the unique spectral features of each element, more detailed structural and electronic information on the bimetallic system can be obtained. The datasets from the two elements in bimetallic SACs complement and corroborate each other, providing a comprehensive picture of the system.

Bimetallic SACs composed of Cu and Au have been widely employed in the photocatalytic reduction of $CO_2$ to $C_2H_4$. However, the detailed photocatalytic mechanism remains ambiguous. To obtain an improved understanding, Xie et al. studied the aggregation and deactivation of CO intermediates during Cu SAC photocatalytic $CO_2$ reduction using HAADF-STEM and XAS. [41] They also successfully synthesized $Cu_5Au_1$ on the surface of $TiO_2$ with isolated Au atoms embedded in the Cu lattice. [42] The existence of Cu-Au bonds was further confirmed by XANES and EXAFS analysis, as shown in Figure 5. Notably, the Au white-line peak in the XANES spectra is relatively weak and insensitive in conventional XAS measurements. To enhance the sensitivity of the Au white-line analysis, we suggest that high-energy resolution fluorescence-detected X-ray absorption spectroscopy (HERFD-XAS) should be employed in future studies. [43]

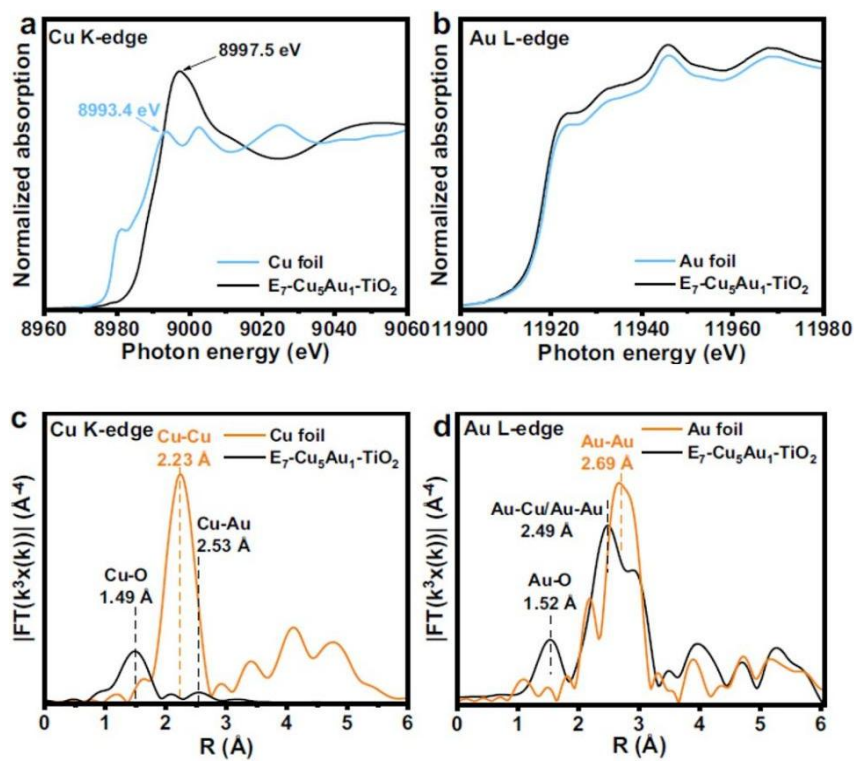

Figure 5: XANES analysis of $Cu_5Au_1$-$TiO_2$ and reference samples at the (a) Cu K-edge and (b) Au $L_3$-edge. Corresponding $K_3$-weighted FT-EXAFS spectra for $Cu_5Au_1$-$TiO_2$ and references at the (c) Cu K-edge and (d) Au $L_3$-edge.[42]

3.2 Bimetallic elements from the same period

For bimetallic SACs consisting of two elements with similar atomic numbers, it is challenging to use conventional structural analysis techniques, such as TEM, to distinguish between the two elements. XAS overcomes this challenge due to its element specific capability. Therefore, it can offer more comprehensive information on the structure and electronic properties of bimetallic SACs from the perspective of each element.

Tang et al. synthesized RuMo/$TiO_2$ bimetallic SACs and conducted a comprehensive XAS analysis to elucidate the structural and electronic properties.[44] The spectra (Figure 6a) and fitting results showed that the valence state of Ru in $Ru_4Mo_1$/$TiO_2$ and $Ru_6$/$TiO_2$ ranged from 0 to +4. Due to electronic interactions, the precense of Mo atoms in $Ru_4Mo_1$/$TiO_2$ caused the Ru atoms to exhibit a higher valence electron density compared to $Ru_6$/$TiO_2$. As the Mo:Ru ratio increased, a significant shift of the Ru absorption edge to higher energy was observed, indicating an increase in the oxidation state of Ru. Conversely, the valence state of Mo in $Ru_4Mo_1$/$TiO_2$ decreased, corresponding to the increase in the Ru valence state, as shown in Figure 6b. This finding supports a strong electronic interaction between Ru and Mo atoms. EXAFS analysis revealed a single peak at ~1.5 Å in Ru$_4$Mo$_1$/TiO$_2$, corresponding to the first shell Ru-O path, as shown in Figure 6c. Similarly, only a single peak at ~1.3 Å attributed to the Mo-O path was observed, as shown in Figure 6d, confirming the atomic dispersion of Mo. These results provide valuable insights into the structural and electronic properties of the RuMo/TiO$_2$ system, highlighting the significant role of XAS in SAC characterization.

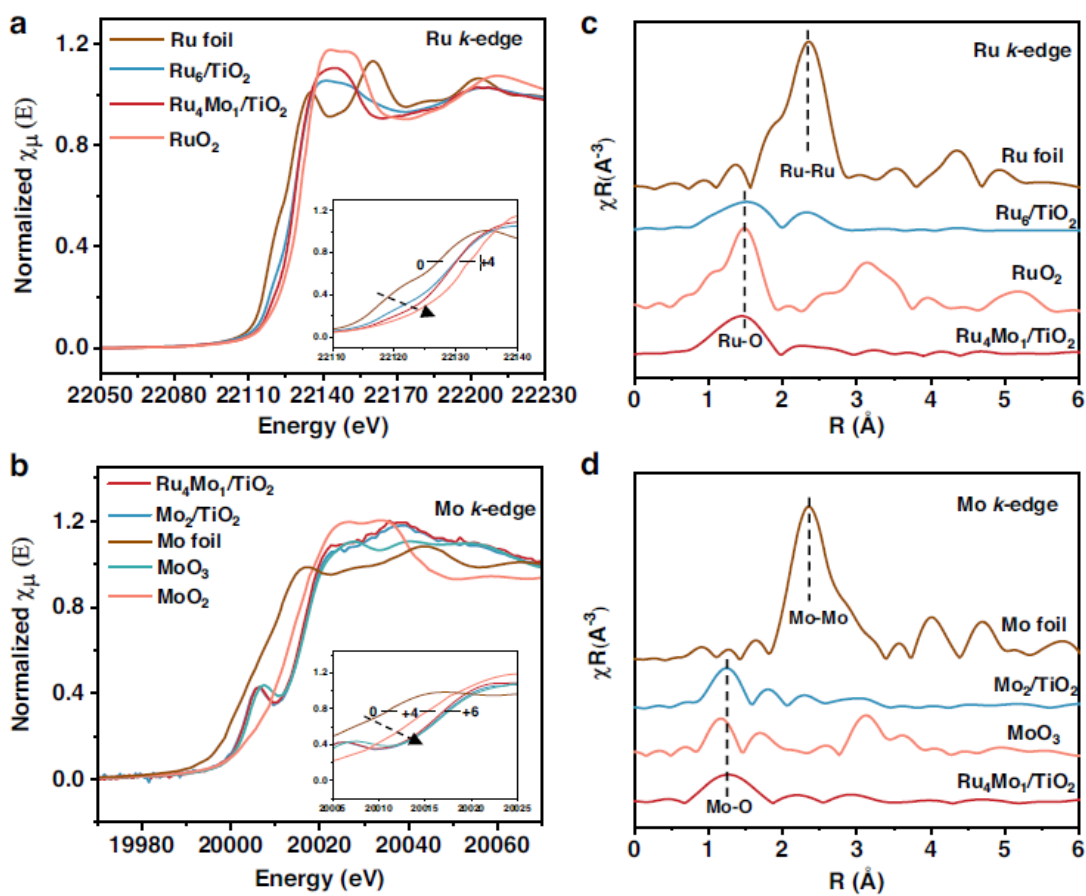

Figure 6: Structural characterization by XAS. Normalized XANES spectra at the (a) Ru K-edge (inset: absorption edge positions of Ru) and (b) Mo K-edge, (inset: absorption edge positions of Mo). FT-EXAFS spectra of (c) Ru-containing samples and (d)Mo-containing samples.[44]

Wei et al. successfully prepared Pt-Au dual-atom catalysts through a two-step deposition−precipitation method.[45] The Pt-Au dual-atom photocatalyst exhibited superior photocatalytic hydrogen production performance compared to monoatomic Pt and Au catalysts. Numerous bright spots were observed in HAADF-STEM images, confirming the distribution of atomically dispersed Au and Pt atoms, as shown in Figures 7a and 7b. However, distinguishing between the two elements is difficult due to their similar atomic numbers. XAS provided element-specific information, revealing the coordination environments of Pt and Au atoms and offering key insights into the active sites of the catalysts. The atomic dispersion and oxidation states were further analyzed by FT-EXAFS spectra (Figures 7c and 7d). These results were consistent with the findings from XPS and HAADF-STEM, enhancing the reliability and comprehensiveness of the structural characterization.

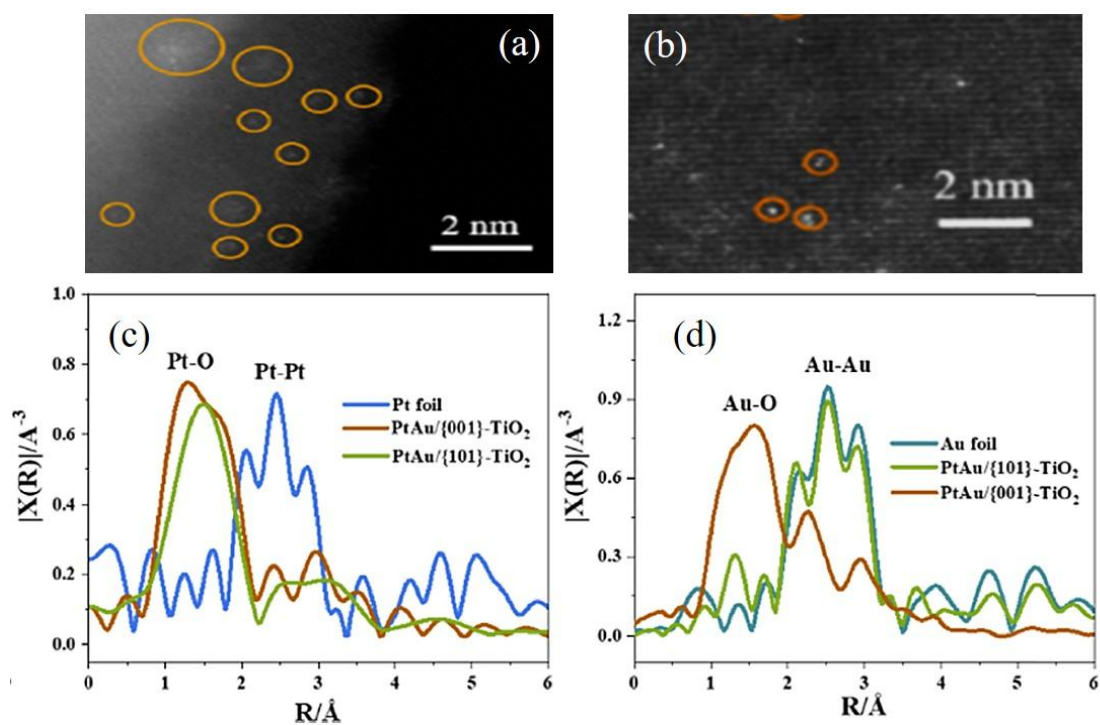

Figure 7: HAADF-STEM images of (a) PtAu/{001}-TiO$_2$ and (b) PtAu/{101}-TiO$_2$. FT-EXAFS spectra of the (c) Pt L$_3$-edge and (d) Au L$_3$-edge.[45]

As Figure 7 shows, when two metals are in the same period, it is difficult to distinguish them at atomic resolution by HAADF-STEM. XAS however remains a robust technique for accurately identifying different metal atoms. The quality of EXAFS analysis depends strongly on the relative positions of the elements in the periodic table. For example, Ru-Mo systems exhibit excellent EXAFS fitting, whereas Pt-Au systems do not. For 4d transition metals, K-edge analysis is typically employed. The K-edge absorption energy difference between Ru and Mo is ~2700 eV, while the $L_3$-edge difference between Pt and Au is only ~360 eV, which is too small to clearly distinguish the two elements. For 5d transition metals, the $L_1$-edge offers higher energy and greater separation of absorption edges for elements with similar atomic numbers. For example, the difference in $L_1$-edge absorption energy is much higher than that of the $L_3$-edge. Therefore, $L_1$-edge analysis is suggested for Au-Pt SACs to improve structural resolution.

4. *In situ* XAS under real conditions

Employing *in situ* techniques enables the real-time monitoring of changes in the surface structure and electronic states of catalysts under real reaction conditions. At present, *quasi-in situ* XPS and *in situ* TEM are predominantly adopted in in situ catalytic research. In comparison, *in situ* XAS possesses two distinct advantages. Firstly, XPS and TEM are typically conducted under vacuum or at least low-pressure conditions. This makes them unsuitable for monitoring photocatalytic reactions at ambient pressure and imposes stringent requirements on reactor design. In contrast, XAS can be performed at ambient pressure, enabling monitoring of photocatalytic reactions under realistic conditions. Secondly, XAS employs high-energy X-rays, which can penetrate the sample cell, making it particularly effective for measurements in liquid-phase or mixed-phase environments. This capability enables real-time tracking of interactions between catalysts and adsorbed reactants during photocatalytic processes, which is crucial for gaining a more complete understanding photocatalytic mechanism.

To better understand catalytic kinetics, Tesvara et al. explored the mechanism of CO oxidation on highly dispersed Pt SACs supported on $TiO_2$ using *in situ* XAS at the Pt $L_3$-edge.[46] By comparing the white-line intensity before and after exposure to CO at 160°C, the oxidation state of the SACs was revealed. Zhou et al. studied ligand-modified Pt SACs supported on $TiO_2$ (Figure 8).[47] *In situ* XAS was performed at different temperatures and under various atmospheres, demonstrating that the activity and stability of the catalysts depended on the choice of ligand and the coordination environment of Pt atoms. Lu et al. reported thermally stable Pd SACs supported on $TiO_2$.[48] The local coordination environment of Pd atoms was monitored in real time by *in situ* XAS under varying atmospheres and temperatures. EXAFS analysis revealed that $H_2$-treated Pd single atoms exhibited decreased Pd-O coordination while forming short Pd-Ti bonds, which stabilized the Pd single atoms. During CO oxidation, Pd-Ti coordination was partially replaced by Pd-O. This unique Pd coordination environment enhanced CO adsorption and activating surrounding oxygen species. This unique coordination environment enabled both Mars－van Krevelen and Eley－Rideal pathways, resulting in excellent catalytic performance. Overall, in situ XAS directly revealed the structural evolution of single-atom sites under different reaction conditions, linking these transformations to kinetic parameters and reaction pathways.

Taken together, these representative studies demonstrate that *in situ* XAS provides detailed, real-time insights into the structural and electronic evolution of SAC photocatalysts. This capability enhances our understanding of photocatalytic mechanisms and guides the rational design of high-performance catalytic systems.

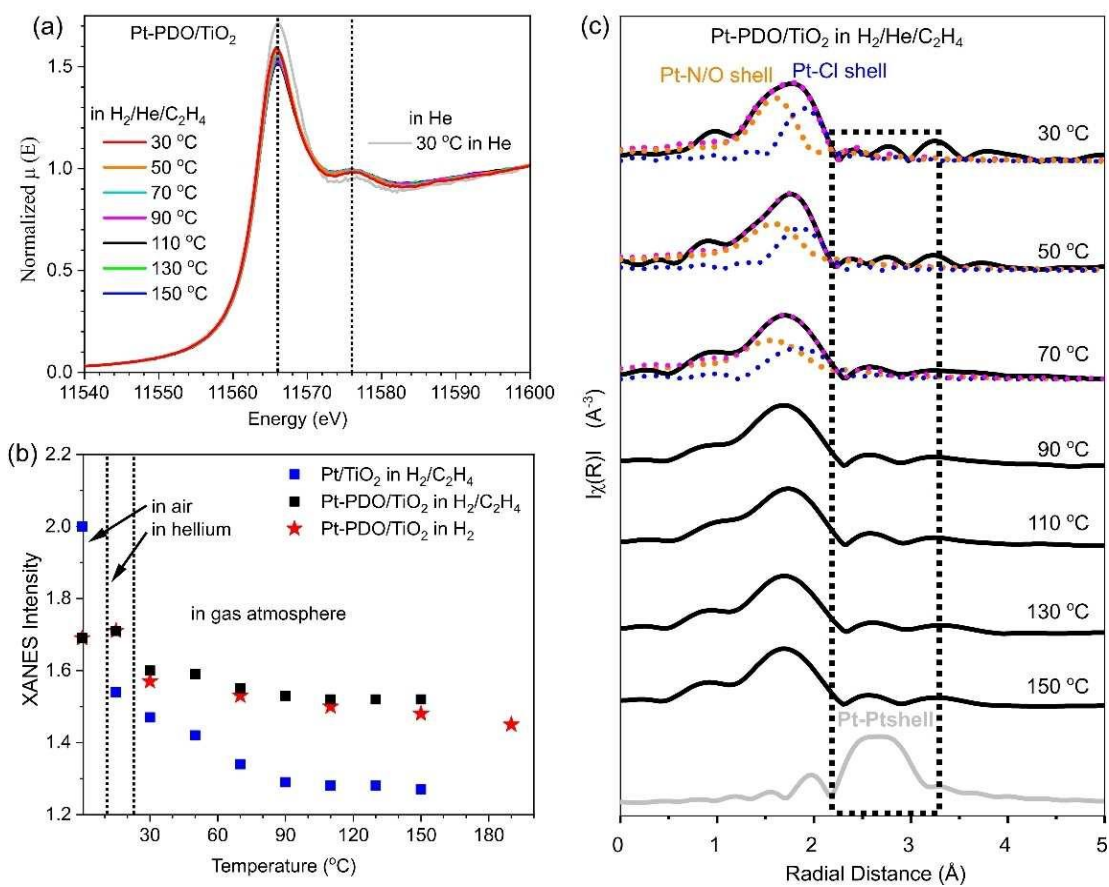

Figure 8: (a) Pt L$_3$-edge XANES spectra of Pt/TiO$_2$ under reaction conditions from 30 to 150°C. (b) XANES white-line intensity for Pt in reaction gas (blue, black) and in H$_2$ (red) at different temperatures. Data points recorded at room temperature in air and at 30°C in helium are shown at the left. (c) Fourier-transformed magnitude of $k^2$-weighted EXAFS spectra of Pt/TiO$_2$ under reaction conditions from 30 to 150°C, with fittings for the first three spectra.[47]

## 5. Conclusion and Outlook

In summary, we have discussed the unique advantages of XAS in characterizing TiO$_2$-supported SACs. For monometallic SACs, XAS provides atomic-scale insights, simultaneously yielding both electronic and structural information. Specifically, XAS analysis elucidates the coordination environment (particularly light-element coordination around metal centers), the oxidation state of metal atoms, and charge transfer mechanisms between SACs and TiO$_2$. For bimetallic SACs, XAS enables element-specific analysis of both electronic and structural properties, providing a comprehensive understanding of the catalysts. By selecting the unique absorption edges of target metals, XAS efficiently distinguishes and identifies metal atoms with similar atomic numbers.

The utilization of high-energy X-rays and the ability to operate without vacuum conditions make XAS particularly advantageous for *in situ* characterization of photocatalytic reactions. It facilitates real-time monitoring of structural evolution under dynamic reaction conditions, including variations in temperature, atmosphere, and liquid-phase environments. This capability overcomes limitations of conventional techniques such as XPS and TEM, providing critical insights into structure−property relationships and valuable guidance for the rational design of high-performance photocatalysts.

Despite these advancements, continued development of XAS is still required. For instance, XANES simulations and EXAFS fitting approaches require further refinement to yield more accurate and detailed information. Additionally, integrating experimental XAS with computational methods such as DFT and machine learning could further enhance its analytical power for studying complex catalytic systems. The application of *in situ* XAS could also be advanced through the design of improved reaction cells and more sensitive X-ray detection equipment.

**Author Contributions:** All the listed authors contributed to the writing of this review manuscript. Writing—original draft preparation and visualization—Y. Li, Writing—review and editing-D. Morris, Conceptualization and writing—review and editing-P. Zhang. All authors have read and agreed to the published version of the manuscript.

**Acknowledgements**: This work was supported by the Research Foundation of Education department of Jilin Province (JJKH20241426KJ) and the Fundamental Research Funds for the Central Universities (2412022QD035). CLS@APS facilities at the Advanced Photon Source (APS) are supported by the U.S. Department of Energy (DOE), NSERC Canada, the University of Washington, the Canadian Light Source (CLS), and the APS. Use of the APS is supported by the DOE under Contract no. DEAC02-06CH11357. The CLS is financially supported by NSERC Canada, CIHR, NRC, and the University of Saskatchewan.

**Data Availability Statement**: Not applicable.
**Conflicts of Interest**: The authors declare no conflict of interest.

# References

1. a) A. Fujishima, X. Zhang, D. A. Tryk, TiO$_2$ photocatalysis and related surface phenomena, *Surf. Sci. Rep.*, **2008**, 63, 515-582, b) Q. Guo, C. Zhou, Z. Ma, X. Yang. Fundamentals of TiO$_2$ photocatalysis: concepts, mechanisms, and challenges. *Adv Mater*, **2019**, 31, 1901997.
2. a) H. Park, Y. Park, W. Kim, W. Choi. Surface modification of TiO$_2$ photocatalyst for environmental applications. *J Photoch Photobio C*, **2013**, 15, 1-20, b) J. Wen,

X. Li, W. Liu, Y. Fang, J. Xie, Y. Xu. Photocatalysis fundamentals and surface modification of $TiO_2$ nanomaterials. *Chin J Catal*, **2015**, 36, 2049-2070.

3. a) W. H. Lee, C. W. Lee, G. D. Cha, B.-H. Lee, J. H. Jeong, H. Park, J. Heo, M. S. Bootharaju, S.-H. Sunwoo, J. H. Kim, K. H. Ahn, D.-H. Kim, T. Hyeon. Floatable photocatalytic hydrogel nanocomposites for large-scale solar hydrogen production. *Nature Nanotech*, **2023**, 18(7), 754-762, b) W. Song, K. C. Chong, G. Qi, Y. Xiao, G. Chen, B. Li, Y. Tang, X. Zhang, Y. Yao, Z, Lin, Z. Zou, B. Liu. Unraveling the transformation from type-II to Z-scheme in perovskite-based heterostructures for enhanced photocatalytic $CO_2$ reduction. *J Am Chem Soc*, **2024**, 146(5), 3303-3314.

4. a) S. Hejazi, S. Mohajernia, B. Osuagwu, G. Zoppellaro, P. Andryskova, O. Tomanec, S. Kment, R. Zbořil, P. Schmuki. *Adv Mater*, **2020**, 32, 1908505, b) J. E. Yoo, K. Lee, M. Altomare, Self-organized arrays of single-metal catalyst particles in $TiO_2$ cavities: a highly efficient photocatalytic system. *Angew Chem Int Ed*, **2013**, 52, 7514 –7517.

5. a) J-C. Liu, Y. Tang, Y-G. Wang, T. Zhang, J. Li. Theoretical understanding of the stability of single-atom catalysts. *Natl Sci Rev*, **2018**, 5, 638-641, b) K. Kruczała, S. Neubert, K. Dhaka, D. Mitoraj, P. Jánošíková, C. Adler, I. Krivtsov, J. Patzsch, J. Bloh, J. Biskupek, U. Kaiser, R. K. Hocking, M. C. Toroker, R. Beranek. Enhancing Photocatalysis: Understanding the Mechanistic Diversity in Photocatalysts Modified with Single-Atom Catalytic Sites. *Adv Sci*, **2023**, 10, 2303571.

6. a) Y. Wang, N. Denisov, S. Qin, D. S. Gonçalves, H. Kim, B. B. Sarma, P. Schmuki. Stable and highly active single atom configurations for photocatalytic $H_2$ generation. *Adv Mater*, **2024**, 36, 2400626, b) Z. Guo, G. Jing, S. A. Tolba, C.-s. Yuan, Y.-H. Li, X. Zhang, Z. Huang, H. Zhao, X. Wu, H. Shen, W. Xia. Design and construction of an O-Au-O coordination environment in Au single atom-doped $Ti^{4+}$ defected $TiO_2$ for an enhanced oxidative ability of lattice oxygen for $Hg^0$ oxidation. *Chem Eng J*, **2023**, 451, 138895, c) K. Fujiwara, S. E. Pratsinis,

Single Pd atoms on TiO$_2$ dominate photocatalytic NO$_x$ removal. *Appl Catal B*, **2018**, 226, 127-134, d) A. H. Jenkins, E. E. Dunphy, M. F. Toney, C. B. Musgrave, J. W. Medlin. Tailoring the near-surface environment of Rh single-atom catalysts for selective CO$_2$ hydrogenation. *ACS Catal*, **2023**, 13, 15340-15350, e) B. H. Lee, S. Park, M. Kim, A. K. Sinha, S. C. Lee, E. Jung, W. J. Chang, K.-S. Lee, J. H. Kim, S.-P. Cho, H. Kim, K. T. Nam, T. Hyeon. Reversible and cooperative photoactivation of single-atom Cu/TiO$_2$ photocatalysts. *Nat mater*, **2019**, 18, 620-626.

7. C. T. Chantler, G. Bunker, P. D'Angelo, S. Diaz-Moreno, X-ray absorption spectroscopy, *Nat Rev Methods Primers*, **2024**, 4, 89.

8. a) S. Duan, R. Wang, J, Liu, Stability investigation of a high number density Pt$_1$/Fe$_2$O$_3$ single-atom catalyst under different gas environments by HAADF-STEM, *Nanotechnology*, **2018**, 29, 204002, b) P. Yang, Z. Li, Y. Yang, R. Li, L. Qin, Y. Zou, Effects of electron microscope parameters and sample thickness on high angle annular dark field imaging, *Scanning*, **2022**, 8503314.

9. Z. Chen, A. G. Walsh, P. Zhang, Structural analysis of single-atom catalysts by X-ray absorption spectroscopy. *Acc Chem Res*, **2024**, 57, 521−532

10. a) A. L. Ankudinov, J. J. Rehr, Development of XAFS theory. *Synchrotron Radia.*, **2003**, 10, 366-368, b) M. Xu, S. He, H. Chen, G. Cui, L. Zheng, B. Wang, M. Wei, TiO$_{2−x}$-modified Ni nanocatalyst with tunable metal–support interaction for water–gas shift reaction. *ACS Catal*, **2017**, 7, 7600-7609, c) P. N. Duchesne, G. Chen, N. Zheng, P. Zhang, Local structure, electronic behavior, and electrocatalytic reactivity of CO-reduced platinum–iron oxide nanoparticles. *J Phys Chem C*, **2013**, 117, 26324−26333, d) Z. Hu, X. Li, S. Zhang, Q. Li, J. Fan, X. Qu, L. Lv, Fe$_1$/TiO$_2$ hollow microspheres: Fe and Ti dual active sites boosting the photocatalytic oxidation of NO. *Small*, **2020**, 16, 2004583.

11. V. Schwartz, D. R. Mullins, W. Yan, H. Zhu, S. Dai, S. H. Overbury. Structural investigation of Au catalysts on TiO$_2$− SiO$_2$ supports: nature of the local structure of Ti and Au atoms by EXAFS and XANES. *J Phy Chem C*, **2007**, 111, 17322-17332.

12. a) B. Han, Y. Guo, Y. Huang, W. Xi, J. Xu, J. Luo, H. Qi, Y. Ren, X. Liu, B. Qiao, T. Zhang. Strong metal–support interactions between Pt single atoms and $TiO_2$. *Angew Chem Int Ed*, **2020**, 59, 11824–11829, b) Y. Guo, Y. Huang, B. Zeng, B. Han, M. AKRI, M. Shi, Y. Zhao, Q. Li, Y. Su, L. Li, Q. Jiang, Y.-T. Cui, L. Li, R. Li, B. Qiao, T. Zhang. Photo-thermo semi-hydrogenation of acetylene on $Pd_1$/$TiO_2$ single-atom catalyst. *Nat comm*, **2022**, 13, 2648, c) D. M. Chevrier, R. Yang, A. Chatt, P. Zhang. Bonding properties of thiolate-protected gold nanoclusters and structural analogs from X-ray absorption spectroscopy. *Nanotechnology Rev*, **2015**, 4, 193-206.

13. a) H. Wang, H. Qi, X. Sun, S. Jia, X. Li, T. J. Miao, L. Xiong, S. Wang, X. Zhang, X. Liu, A. Wang, T. Zhang, W. Huang, J. Tang. High quantum efficiency of hydrogen production from methanol aqueous solution with PtCu–$TiO_2$ photocatalysts. *Nat Mater*, **2023**, 22, 619-626, b) C. Wang, K. Wang, Y. Feng, C. Li, X. Zhou, L. Gan, Y. Feng, H. Zhou, B. Zhang, X. Qu, H. Li, J. Li, A. Li, Y. Sun, S. Zhang, Guo Yang, Y. Guo, S. Yang, T. Zhou, F. Dong, K. Zheng, L. Wang, J. Huang, Z. Zhang, X. Han. Co and Pt dual-single-atoms with oxygen-coordinated Co–O–Pt dimer sites for ultrahigh photocatalytic hydrogen evolution efficiency. *Adv Mater*, **2021**, 33, 2003327.

14. P. Liu, Y. Zhao, R. Qin, P. Liu, Y. Zhao, R. Qin, S. Mo, Gu. Chen, L. Gu, D. M. Chevrier, P. Zhang, Q. Guo, D. Zang, B. Wu, G. Fu, N. Zheng. Photochemical route for synthesizing atomically dispersed palladium catalysts. *Science*, **2016**, 352, 797-800.

15. a) M. Xiao, L. Zhang, B. Luo, M. Lyu, Z. Wang, H. Huang, S. Wang, A. Du, L. Wang. Molten-salt-mediated synthesis of an atomic nickel Co-catalyst on $TiO_2$ for improved photocatalytic $H_2$ evolution. *Angew Chem Int Ed*, **2020**, 59, 7230–7234, b) J. Chen, Y. Kang, W. Zhang, Z. Zhang, Y. Chen, Y. Yang, L. Duan, Y. Li, W. Li. Lattice-confined single-atom $Fe_1S_x$ on mesoporous $TiO_2$ for boosting ambient electrocatalytic $N_2$ reduction reaction. *Angew Chem Int Ed*, **2022**, 134, e202203022, c) M. Yang, H. Li, F. Liu, S. Sun, J. Mei, Y. Jiao, J. Cui, Y. Xu, H. Song, Z. Duan, W. Liu, Y. Ren. Mechanism insight into oxygen vacancy-dependent effect in $Fe_1/TiO_2$ single-atom catalyst for highly enhanced photo-Fenton mineralization of phenol. *Appl Catal B*, **2024**, 354, 124071.

16. S. R. Bare, T. Ressler, Characterization of catalysts in reactive atmospheres by X-ray absorption spectroscopy. *Adv catal*, **2009**, 52, 339-465.

17. a) X. Li, X. Yang, J. Zhang, Y. Huang, B. Liu. In situ/operando techniques for characterization of single-atom catalysts. *ACS Catal*, **2019**, 9, 2521-2531, b) L. Chen, S. I. Allec, M. T. Nguyen, L. Kovarik, A. S. Hoffman, J. Hong, D. Meira, H. Shi, S. R. Bare, V.-A. Glezakou, R. Rousseau, J. Szanyi. Dynamic evolution of palladium single atoms on anatase titania support determines the reverse water–gas shift activity. *J Am Chem Soc*, **2023**, 145, 10847-10860, c) N. C. Nelson, L. Chen, D. Meira, L. Kovarik, J. Szanyi. In situ dispersion of palladium on $TiO_2$ during reverse water–gas shift reaction: formation of atomically dispersed palladium. *Angew Chem Int Ed*, **2020**, 132, 17810-17816.

18. P. D. Robb, M. Finnie, P. Longo, A. J. Craven, Experimental evaluation of interfaces using atomic-resolution high angle annular dark field (HAADF) imaging, *Ultramicroscopy*, **2022**, 114, 11-19.

19. S. Kim, E. Sasmaz, Transformation of Platinum Single-Atom Catalysts to Single-Atom Alloys on Supported Nickel: TEM and XAS Spectroscopic Investigation. *ChemCatChem*, **2022**, 14, e202200568.

20. J. Chaboy, J. García, A. Marcelli, M.R. Ruiz-López. An experimental and theoretical study of multi-electron excitations at the $L_3$ absorption edge in some rare earth alloys and their hydrides. *Chem phy letters*, **1990**, 174, 389-395.

21. H. Shin, W.-G. Jung, D.-H. Kim, J.-S. Jang, Y. H. Kim, W.-T. Koo, J. Bae, C. Park, S.-H. Cho, B. J. Kim, I.-D. Kim, Single-Atom Pt Stabilized on One Dimensional Nanostructure Support via Carbon Nitride/$SnO_2$ Heterojunction Trapping. *ACS Nano*, **2020**, 14, 11394-11405.

22. Z. H.N. Al-Azri, W.-T. Chen, A. Chan, V. Jovic, T. Ina, H. Idriss, G. I.N. Waterhouse, The roles of metal co-catalysts and reaction media in photocatalytic hydrogen production: Performance evaluation of M/TiO$_2$ photocatalysts (M=Pd, Pt, Au) in different alcohol–water mixtures. *J Catal*, **2015**, 329, 355-367.

23. Y. Chen, S. Ji, W. Sun, Y. Lei, Q. Wang, A. Li, W. Chen, G. Zhou, Z. Zhang, Y. Wang, L. Zheng, Q. Zhang, L. Gu, X. Han, D. Wang, Y. Li. Engineering the atomic interface with single platinum atoms for enhanced photocatalytic hydrogen production. *Angew Chem Int Ed,* **2020**, 132, 1311-1317.

24. Y. Feng, C. Wang, P. Cui, C. Li, B. Zhang, L. Gan, S. Zhang, X. Zhang, X. Zhou, Z. Sun, K. Wang, Y. Duan, H. Li, K. Zhou, H. Huang, A. Li, C. Zhuang, L. Wang, Z. Zhang, X. Han. Ultrahigh photocatalytic CO$_2$ reduction efficiency and selectivity manipulation by single-tungsten-atom oxide at the atomic step of TiO$_2$. *Adv Mater*, **2022**, 34, 2109074.

25. Y. Li, A. G. Walsh, D. Li, D. Do, H. Ma, C. Wang, P. Zhang, X. Zhang. W-Doped TiO$_2$ for photothermocatalytic CO$_2$ reduction. *Nanoscale*, **2020**, 12, 17245-17252.

26. M. Guo, Q. Meng, M. L. Gao, L. Zheng, Q. Li, L. Jiao, H.-L. Jiang. Single-Atom Pt Loaded on MOF-Derived Porous TiO$_2$ with Maxim-Ized Pt Atom Utilization for Selective Hydrogenation of Halonitro-benzene. *Angew Chem Int Ed*, **2024**, e202418964.

27. Q. He, A. Li, B. Yao, W. Zhou, C. J. Kiely, STEM High Angle Annular Dark-Field Imaging, *Springer Handbook of Advanced Catalyst Characterization*, **2023**, 409–448.

28. C. Wang, Z. Wang, S. Mao, Z. Chen, Y. Wang. Coordination environment of active sites and their effect on catalytic performance of heterogeneous catalysts, *Chinese J Catal*, 2022, 43, 928-955.

29. a) D. Moonshiram, Y. Pineda-Galvan, D. Erdman, M. Palenik, R. Zong, R. Thummel, Yulia Pushkar, Uncovering the Role of Oxygen Atom Transfer in Ru-Based Catalytic Water Oxidation, *J Am Chem Soc*, **2016**, 138, 15605−15616, b) M. A. Soldatov, A. Martini, A. L. Bugaev, I. Pankin, P. V. Medvedev, A. A. Guda, A. M. Aboraia, Y. S. Podkovyrina, A. P. Budnyk, A. A. Soldatov, C. Lamberti, The insights from X-ray absorption spectroscopy into the local atomic structure and chemical bonding of Metal–organic frameworks, *Polyhedron*, **2018**, 155, 232-253.

30. Y. Guo, Y. Huang, B. Zeng, B. Han, M. AKRI, M. Shi, Y. Zhao, Q. Li, Y. Su, L. Li, Q. Jiang, Y.-T. Cui, L. Li, R. Li, B. Qiao, T. Zhang. Photo-thermo semi-hydrogenation of acetylene on $Pd_1/TiO_2$ single-atom catalyst. *Nat comm*, **2022**, 13, 2648.

31. a) J. Xu, X. Su, H. Duan, B. Hou, Q. Lin, X. Liu, X. Pan, G. Pei, H. Geng, Y. Huang, T. Zhang. Influence of pretreatment temperature on catalytic performance of rutile $TiO_2$-supported ruthenium catalyst in $CO_2$ methanation, *J Catal*, **2016**, 333, 227-237, b) T. A. Le, Y. Kim, H. W. Kim, S.-U. Lee, J.-R. Kim, T.-W. Kim, Y.-J. Lee, H.-J. Cha, *Appl Catal B*, **2021**, 285, 119831.

32. A. Vojvodic, J. K. Nørskov, F. Abild-Pedersen, Electronic Structure Effects in Transition Metal Surface Chemistry, *Top Catal*, **2014**, 57, 25–32.

33. Z. Shi, J. Li, Y. Wang, S. Liu, J. Zhu, J. Yang, X. Wang, J. Ni, Z. Jiang, L. Zhang, Y. Wang, C. Liu, W. Xing, J. Ge. Customized reaction route for ruthenium oxide towards stabilized water oxidation in high-performance PEM electrolyzers. *Nat Comm*, **2023**, 14, 843.

34. I. Alperovich, G. Smolentsev, D. Moonshiram, J. W. Jurss, J. J. Concepcion, T. J. Meyer, A. Soldatov, Y. Pushkar. Understanding the electronic structure of 4d metal complexes: from molecular spinors to L-edge spectra of a di-Ru catalyst. *J Am Chem Soc*, **2011**, 133, 15786-15794.

35. H. A. Suarez Orduz, L. Bugarin, S.-L. Heck, P. Dolcet, M. Casapu, J.-D. Grunwaldt, P. Glatzel. $L_3$-edge X-ray spectroscopy of rhodium and palladium compounds. *J Synchrotron Radia*, **2024**, 31, 733-740.

36. X. Ruan, S. Li, C. Huang, W. Zheng, X. Cui, S. K. Ravi. Catalyzing Artificial Photosynthesis with $TiO_2$ Heterostructures and Hybrids: Emerging Trends in a Classical yet Contemporary Photocatalyst. *Adv Mater*, **2024**, 36, 2305285.

37. a) H. Sang Jeon, J. Timoshenko, F. Scholten, I. Sinev, A. Herzog, F. T. Haase, B. Roldan Cuenya. Operando Insight into the Correlation between the Structure and

Composition of CuZn Nanoparticles and Their Selectivity for the Electrochemical $CO_2$ Reduction. *J Am Chem Soc.* **2019**, 141, 19879−19887, b) D. Lützenkirchen-Hecht, M. Wagemaker, P. Keil, A. A. van Well, R. Frahm, Ex situ reflection mode EXAFS at the Ti K-edge of lithium intercalated $TiO_2$ rutile, *Surf Sci,* **2003**, 538, 10–22.

38. Y. Fang, Q. Zhang, H. Zhang, X. Li, W. Chen, J. Xu, H. Shen, J. Yang, C. Pan, Y. Zhu, J. Wang, Z. Luo, L. Wang, X. Bai, F. Song, L. Zhang, Y. Guo. Dual activation of molecular oxygen and surface lattice oxygen in single atom $Cu_1/TiO_2$ catalyst for CO oxidation. *Angew Chem Int Ed*, **2022**, 61, e202212273.

39. Y. Li, Z.-S. Wu, P. Lu, X. Wang, W. Liu, Z. Liu, J. Ma, W. Ren, Z. Jiang, X. Bao. High-Valence Nickel Single-Atom Catalysts Coordinated to Oxygen Sites for Extraordinarily Activating Oxygen Evolution Reaction. *Adv Sci*, **2020**, 7, 1903089.

40. Y. Shen, C. Ren, L. Zheng, X. Xu, R. Long, W. Zhang, Y. Yang, Y. Zhang, Y. Yao, H. Chi, J. Wang, Q. Shen, Y. Xiong, Z. Zou, Y. Zhou. Room-temperature photosynthesis of propane from $CO_2$ with Cu single atoms on vacancy-rich $TiO_2$. *Nat comm*, **2023**, 14,1117.

41. Z. Xie, L. Li, S. Gong, S. Xu, H. Luo, D. Li, H. Chen, M. Chen, K. Liu, W. Shi, D. Xu, Y. Lei. *Angew. Chem. Int. Ed.* **2024**, *63*, e202410250.

42. Z. Xie, S. Xu, L. Li, S. Gong, X. Wu, D. Xu, B. Mao, T. Zhou, M. Chen, X. Wang, W. Shi, S. Song. Well-defined diatomic catalysis for photosynthesis of $C_2H_4$ from $CO_2$. *Nat comm*, **2024**, 15, 2422.

43. M. Bauer. HERFD-XAS and valence-to-core-XES: new tools to push the limits in research with hard X-rays? *Phys Chem Chem Phys*, **2014**, 16, 13827-13837.

44. M. Tang, J. Shen, Y. Wang, Y. Zhao, T. Gan, X. Zheng, D. Wang, B. Han, Z. Liu. Highly efficient recycling of polyester wastes to diols using Ru and Mo dual-atom catalyst. *Nat comm*, **2024**, 15, 5630.

45. T. Wei, P. Ding, T. Wang, L.-M. Liu, X. An, X. Yu. Facet-regulating local coordination of dual-atom cocatalyzed $TiO_2$ for photocatalytic water splitting. *ACS Catal*, **2021**, 11, 14669-14676.

46. C. Tesvara, M. R. Yousuf, M. Albrahim, D. Troya, A. Shrotri, E. Stavitski, A. M. Karim, P. Sautet. Unraveling the CO Oxidation Mechanism over Highly Dispersed Pt Single Atom on Anatase TiO$_2$ (101). *ACS Catal*, **2024**, 14, 7562-7575.

47. X. Zhou, G. E. Sterbinsky, E. Wasim, L. Chen, S. L. Tait. Tuning Ligand-Coordinated Single Metal Atoms on TiO$_2$ and their Dynamic Response during Hydrogenation Catalysis. *ChemSusChem*, **2021**, 14, 3825-3837.

48. Y. Lu, F. Lin, Z. Zhang, C. Thompson, Y. Zhu, N. Doudin, L. Kovarik, C. E. G. Vargas, D. Jiang, J. L. Fulton, Y. Wu, F. Gao, Z. Dohnálek, A. M. Karim, H. Wang, Y. Wang. Enhancing activity and stability of Pd-on-TiO$_2$ single-atom catalyst for low-temperature CO oxidation through in situ local environment tailoring. *J Am Chem Soc*, **2024**, 146, 28141-28152.